\documentclass[12pt]{iopart}

\usepackage{graphicx}

\def\eion{{(e~+~ion)}\ }

\def\fexvii{{\rm Fe~\sc xvii}\ }
\def\fexviii{{\rm Fe~\sc xviii}\ }

\def\en{{$n$\ }}
\def\el{{$l$\ }}
\def\ii{{$i$\ }}
\def\dne{{$N_e$\ }}
\def\om{{$\omega$\ }}
\def\sig{{$\sigma$\ }}
\def\gam{{$\Gamma$\ }}
\def\gamc{{$\Gamma_c$\ }}
\def\gamd{{$\Gamma_d$\ }}

\newcommand{\be}{\begin{equation}}
\newcommand{\ee}{\end{equation}}
\begin{document}

\title[Plasma broadening of resonances]{Plasma broadening
of autoionizing resonances}

\author{A K Pradhan$^{1,2,3}$}

\address{$^1$ Department of Astronomy, $^2$ Chemical Physics Program, 
$^3$ Biophysics Graduate Program,
Ohio State University, Columbus, Ohio 43210, USA}
\vspace{10pt}
\begin{indented}
\item[]September 2022
\end{indented}

\begin{abstract}
A general formulation is developed to demonstrate
 that atomic autoionizing (AI) resonances are
broadened and shifted significantly due to plasma effects
across bound-free continua. The theoretical and
computational method presented
accounts for broadening mechanisms: electron collisional, ion
microfields (Stark), thermal Doppler, core
excitations, and free-free transitions. {\it Extrinsic} plasma broadening
redistributes and shifts AI resonance strengths while broadly preserving
naturally {\it intrinsic} asymmetries of resonance profiles.
Integrated oscillator strengths are conserved as
resonance structures dissolve into continua with increasing electron density.
As exemplar, the plasma attenuation of photoionization cross sections 
computed using the R-matrix method is studied in neon-like Fe~XVII
in a critical range $N_e = 10^{21-24}$cc along isotherms 
$T = 1-2 \times 10^6$K, and its impact on Rosseland Mean opacities. The 
energy-temperature-density dependent
cross sections would elicit and introduce physical features in resonant
processes in photoionization, \eion excitation and recombination.
The method should be generally applicable to
atomic species in high-energy-density (HED)
sources such as fusion plasmas and stellar interiors.
\end{abstract}

%
%
%
%
%

\section{Introduction}
 Autoionization (AI) resonances are ubiquitous in atomic reactions. As
bound-quasibound-free transitions they
manifest themselves in cross sections of atomic processes in a wider
variety of shapes, sizes and extent in energy than line formation due to
bound-bound transitions. As such, due to generally higher
{\it intrinsic} decay rates
relative to radiative rates, it would be expected that AI resonances
broaden, smear out, and dissolve into the bound-free continuum far more
readily than lines when subjected to {\it extrinsic} high-energy-density
(HED) plasma environments. However, whereas line broadening is an
advanced field and is accounted for via elaborate
treatments,
a theoretical method for AI resonance broadening is not available.
Also, although line broadening treatments are precise for hydrogenic and
simple atomic systems, several approximations need to be made using
fitting
formulae, Gaunt factors, etc., for complex atomic systems involving
large numbers of transitions for practical applications in laboratory
and astrophysical plasmas
\cite{b11,d06,hm15,op95,aas,p81,dk81,dk87,k99}.

 Whereas the main
broadening mechanisms in AI broadening are physically similar to line
broadening, their
theoretical and computational treatment is quite different. Superimposed
on intrinsic AI broadening in atomic cross sections
the extent of resonances owing to extrinsic plasma
effects renders much of the line broadening theory inapplicable,
particularly for multi-electron systems. The {\it unbroadened}
AI resonances themselves vary by orders of magnitude in width, shapes
and heights, and
incorporate two types: large features due to
photoexcitation-of-core (PEC) below thresholds corresponding
to dipole core transitions \cite{ys87}, and infinite
Rydberg series of
resonances converging on to each excited core level of the \eion
system. The generally employed Voigt line profiles obtained
by convolution of a Lorentzian function for radiative and
collisional broadening, and a Gaussian function for Doppler or thermal
broadening, are found to be practically inapplicable
for AI broadening. Numerically, the Voigt kernel is ill-conditioned
since the collisional-to-Doppler width ratio \gamc/\gamd
varies over a far wider range
for resonances than lines and therefore unconstrained {\it a priori}
\cite{h65,nb1}.

\section{Theoretical and computational method}
 The physical processes for broadening of AI resonances differ from
lines qualitatively and quantitatively.
However, line broadening processes and formulae may be generalized
to develop a
theoretical treatment and computational algorithm outlined herein
(details to be presented elsewhere).
The convolved bound-free photoionization cross section of level \ii
may be written as:

\be \sigma_i(\omega) = \int \tilde{\sigma}(\omega') \phi
(\omega',\omega) d\omega', \ee

where \sig and $\tilde{\sigma}$ are the cross sections with
plasma-broadened
and unbroadened AI resonance structures, \om is the photon energy
(Rydberg atomic units are used throughout), and $\phi (\omega',\omega)$
 is the normalized 
Lorentzian profile factor in terms of the {\it total} width \gam due to
all
AI broadening processes included:

\be \phi (\omega',\omega) = \frac{\Gamma(\omega)/\pi}{x^2+\Gamma^2}, \ee

where $x \equiv \omega-\omega'$. The crucial difference with line
broadening is
that AI resonances in the \eion system correspond to and are due to
quantum mechanical interference between discretized continua
defined by excited core ion levels in a multitude of channels. The
coupled channel (CC) approximation, such as implemented by the R-matrix
(RM) method
(viz. \cite{b11,op95,aas}), accounts for AI resonances in an \eion
system
with generally asymmetric profiles (unlike line profiles that are
usually symmetric).
 Schematically, CC-RM calculations are as shown in Fig.~1
for an \eion system with bound and continuum levels.

\begin{figure}
\begin{center}
{ \includegraphics[width=80mm,height=65mm]{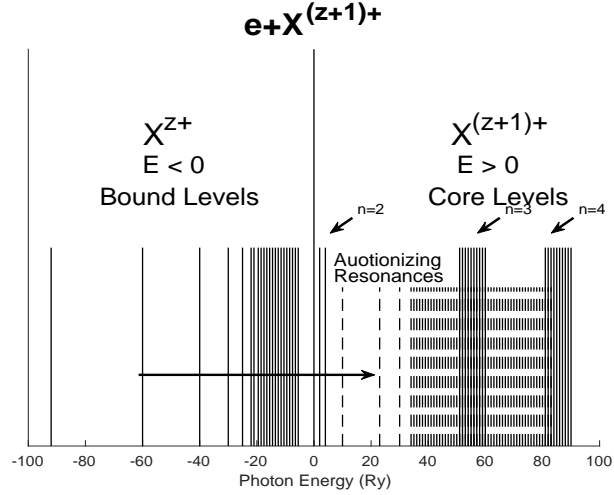}}
\end{center}
\caption{Schematic diagram of coupled channel calculations:
photoionization of bound \eion levels into continua
of core ion levels (solid lines)
$\hbar \omega + X^{z+} \rightarrow e + X^{z+1}$. Rydberg series of
AI resonances (dashed lines) converge on to excited core levels
at energies $E=E_i-z^2/\nu_i^2$, where
$\nu_i$ is a continuous variable corresponding to each threshold energy
$E_i$. The schematics approximately refers to \fexvii cross sections
exemplified in Figs.~2 and 3, with 218 n=2,3,4 energy levels of
\fexviii.
\label{fig:aubro}}
\end{figure}

Given $N$ core ion levels corresponding to resonance
structures,

\be \sigma(\omega) = \sum_i^N \left[ \int \tilde{\sigma}(\omega')
\left[ \frac{\Gamma_i(\omega)/\pi}{x^2 +
\Gamma_i^(\omega)}\right] d \omega' \right]
. \ee

 With $x \equiv \omega' - \omega $, the summation is over all excited
thresholds $E_i$ included in
the $N$-level CC or RM wavefunction expansion, and corresponding
to total damping width $\Gamma_i$ due to all broadening processes.
The profile $\phi(\omega',\omega)$ is centered at each
continuum energy $\omega$, convolved over the variable $\omega'$ and
relative to each excited core ion threshold \ii.
In the present formulation we associate the energy to the effective
quantum number relative to each threshold $\omega' \rightarrow \nu_i$ to
write the total width as:

\begin{eqnarray}
\Gamma_i(\omega,\nu,T,N_e) &  = & \Gamma_c(i,\nu,\nu_c)+
\Gamma_s(\nu_i,\nu_s^*)\\
 & + &  \Gamma_d(A,\omega) + \Gamma_f(f-f;\nu_i,\nu_i'), \nonumber 
\end{eqnarray}

pertaining to
collisional $\Gamma_c$, Stark $\Gamma_s$, Doppler $\Gamma_d$, and
free-free transition $\Gamma_f$ widths
respectively, with additional parameters as defined below.
Without loss of generality we assume a Lorentizan profile
factor that describes collisional-ion broadening which dominates in
HED plasmas. We assume this approximation to be valid since
collisional profile wings extend much wider as $x^{-2}$, compared to
the shorter range $exp(-x^2)$ for thermal Doppler, and $x^{-5/2}$ for
Stark
broadening (viz. \cite{adoc13}). In Eq. (4) the limits $\mp \infty$ are
then replaced by $\mp \Gamma_i/\sqrt{\delta}$; $\delta$ is
chosen to ensure the Lorentzian profile energy range
for accurate normalization.
Convolution by evaluation of Eqs. (1-3) is carried out for each
energy $\omega$ throughout the tabulated mesh of energies used to
delineate all AI resonance structures, for each cross section,
and each core ion threshold. We employ the following expressions for
computations:

\be \Gamma_c(i,\nu) \  = \ 5 \left( \frac{\pi}{kT} \right)^{1/2}
 a_o^3 N_e G(T,z,\nu_i) (\nu_i^4/z^2), \ee

where T, \dne, $z$, and $A$ are the temperature, electron density, ion
charge and atomic weight respectively, and $\nu_i$ is the effective
quantum
number relative to each core ion threshold \ii: $\omega \equiv E =
E_i-\nu_i^2/z^2$ is a continuous variable. The Gaunt factor
$G(T,z,\nu_i) = \sqrt 3/\pi [1/2+ln(\nu_i kT/z)]$
(\cite{adoc13,dk81,x12,nb2}).
A factor
$(n_x/n_g)^4$ is introduced for $\Gamma_c$
to allow for doubly excited AI levels with excited core
evels $n_x$ relative to the ground configuration $n_g$
(e.g. for \fexviii
$n_x=3,4$ relative to the ground configuration $n_g=2$).
A treatment of the Stark effect for complex
systems entails two approaches, one where both electron and ion
perturbations are combined (viz. \cite{dk87,x12}), or separately (viz.
\cite{op95,adoc13}) employed herein. Excited Rydberg levels are nearly
hydrogenic
and ion perturbations are the main broadening effect, though collisional
broadening competes significantly increasing with density
as well as $\nu_i^4$ (Eq.~5).
The total Stark width of a given
\en-complex is $\approx (3F/z)n^2$, where F is the plasma electric
microfield.
Assuming the dominant ion perturbers to be protons and density
equal to electrons, \dne=$N_p$, we take $F=[(4/3)
\pi a_o^3 N_e)]^{2/3}$, as employed in the Mihalas-Hummer-D\"{a}ppen
equation-of-state formulation \cite{mhd}.

\be \Gamma_s(\nu_i,\nu_s^*) =
[(4/3)\pi a_o^3 N_e]^{2/3} \nu_i^2. \ee

In addition, in employing Eq. (6) a Stark ionization parameter
$\nu_s^* = 1.2\times 10^3 N_e^{-2/15}z^{3/5}$ is introduced such
that AI resonances may be considered fully dissolved into the continuum
for $\nu_i > \nu_s^*$ (analogous 
to the Inglis-Teller series limit \cite{it39,mhd}). 
Calculations are carried out with and without
$\nu_s^*$ as shown later in Table~1. The Doppler width is:

\be \Gamma_d (A,T,\omega) = 4.2858 \times 10^{-7} \sqrt(T/A), \ee

where $\omega$ is {\em not} the usual line center but taken to be each
AI resonance energy. The last term $\Gamma_f$ in Eq. (5) accounts for
free-free
transitions among autoionizing levels with $\nu_i,\nu_i'$ such that 

\be X_i + e(E_i,\nu_i) \longrightarrow X_i' + e'(E_i',\nu_i'). \ee

The large number of free-free transition probabilities for $+ve$ energy
AI
levels $E_i,E_i' > 0$ may be computed using RM or atomic structure
codes (viz. \cite{s00,ss}). 

 Whereas Eq.(3) has an analytical solution in terms of
$tan^{-1}(x/\Gamma)/\Gamma$ evaluated at limiting values of $x
\rightarrow \mp \Gamma/\sqrt\delta$, its evaluation
for practical applications entails piece-wise integration across
multiple energy ranges spanning many excited thresholds and different
boundary
conditions. For example, the total width $\Gamma$ is very large at
high densities and the Lorentzian profile may be incomplete
above the ionization threshold and therefore not properly normalized.
We obtain the necessary redward left-wing correction for partial
renormalization as

\be
\lim_{a \rightarrow - \Gamma/2\sqrt\delta}
\int_a^{+\Gamma/\sqrt\delta} \phi(\omega,\omega') d\omega' =
\left[ \frac{1}{4} - \frac{tan^{-1}(\frac{a}{\Gamma/2\sqrt\delta})}{\pi}
\right],
\ee

where $a$ is the lower energy range up to the ionization threshold,
reaching
the maximum value $-\Gamma/2\sqrt\delta$.

\section{Results and discussion}
The complexity and magnitude of computations is demonstrated
for \fexvii that is of considerable importance in astrophysical and
laboratory
plasmas described in a number of previous works (\cite{np16} and
references therein), owing to its neon-like ground configuration
and many excited configurations and $n$-complexes of levels
and transitions. We utilize new results from an extensive
Breit-Pauli R-Matrix (BPRM) calculation
with 218 fine structure levels dominated by $n=2,3,4$ levels of the core
ion \fexviii (to be reported elsewhere). The 587 \fexvii
bound levels ($E<0$) considered
are dominated by configurations $1s^22s^22p^6 (^1S_0),
1s^22s^p2p^qn\ell, [SLJ] \ (p,q
= 0-2, \ n \leq 10, \ \ell \leq 9, \ J \leq 12$). The core \fexvii
levels
included in the CC calculation for the (e~+~\fexviii) $\rightarrow
$\fexvii system
are:$1s^22s^22p^5 (^2P^o_{1/2,3/2}), 1s^22s^22p^q,n\ell, [S_iL_iJ_i] \
(p=4,5, \ n \leq 4, \ell \leq 3)$. The Rydberg series of
AI resonances correspond to $(S_iL_iJ_i) \ n \ell, \ n \leq 10, \ell
\leq 9$, with effective quantum number defined as a continuous variable
$\nu_i = z/\sqrt(E_i-E) \ (E>0)$, throughout the energy range up to the
highest
$218^{th}$ \fexviii core level; the $n=2,3,4$ core levels range from
E=0-90.7 Ry (\cite{n11,np16}).
 The \fexvii BPRM calculations were carried out resolving the bound-free
cross sections at $\sim$40,000 energies for 454
bound levels with AI resonance structures.
Given 217 excited core levels of \fexviii, convolution
is carried out at each energy or approximately $10^9$ times for each
(T,$N_e$) pair.

\begin{figure}
\begin{center}
{ \includegraphics[width=80mm,height=110mm]{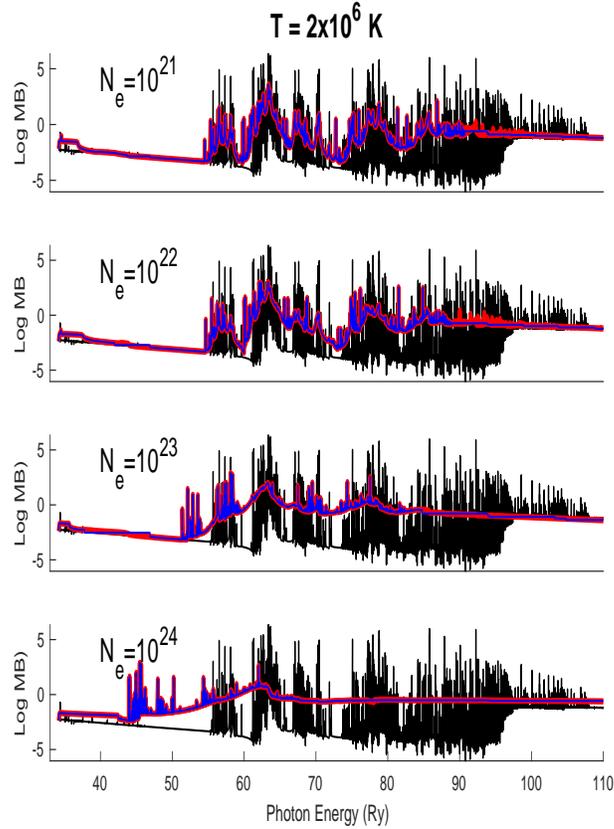}}
\end{center}
\vskip 0.25in
\caption{Plasma broadened photoionization cross sections
for $\hbar \omega + \fexvii \rightarrow e~+~\fexviii $ of the bound
level $2s^22p^5[^2P^o_{3/2}]3p (^3D_2)$ (ionization energy 37.707
Ry) along isotherm
$T=2 \times 10^6$K and electron densities $N_e=10^{21,22,23}$cc: black
---
unbroadened, red --- broadened, blue --- broadened with Stark
ionization cut-off $\nu_s^*$ (Table 1). Rydberg series of
AI resonance complexes with $\nu_i \leq 10$ belonging to 217 excited
\fexviii levels $E_i$ broaden and shift with increasing density, also
 resulting in continuum raising and threshold lowering.
\label{fig:t63}}
\end{figure}

\begin{figure}
\begin{center}
{ \includegraphics[width=80mm,height=90mm]{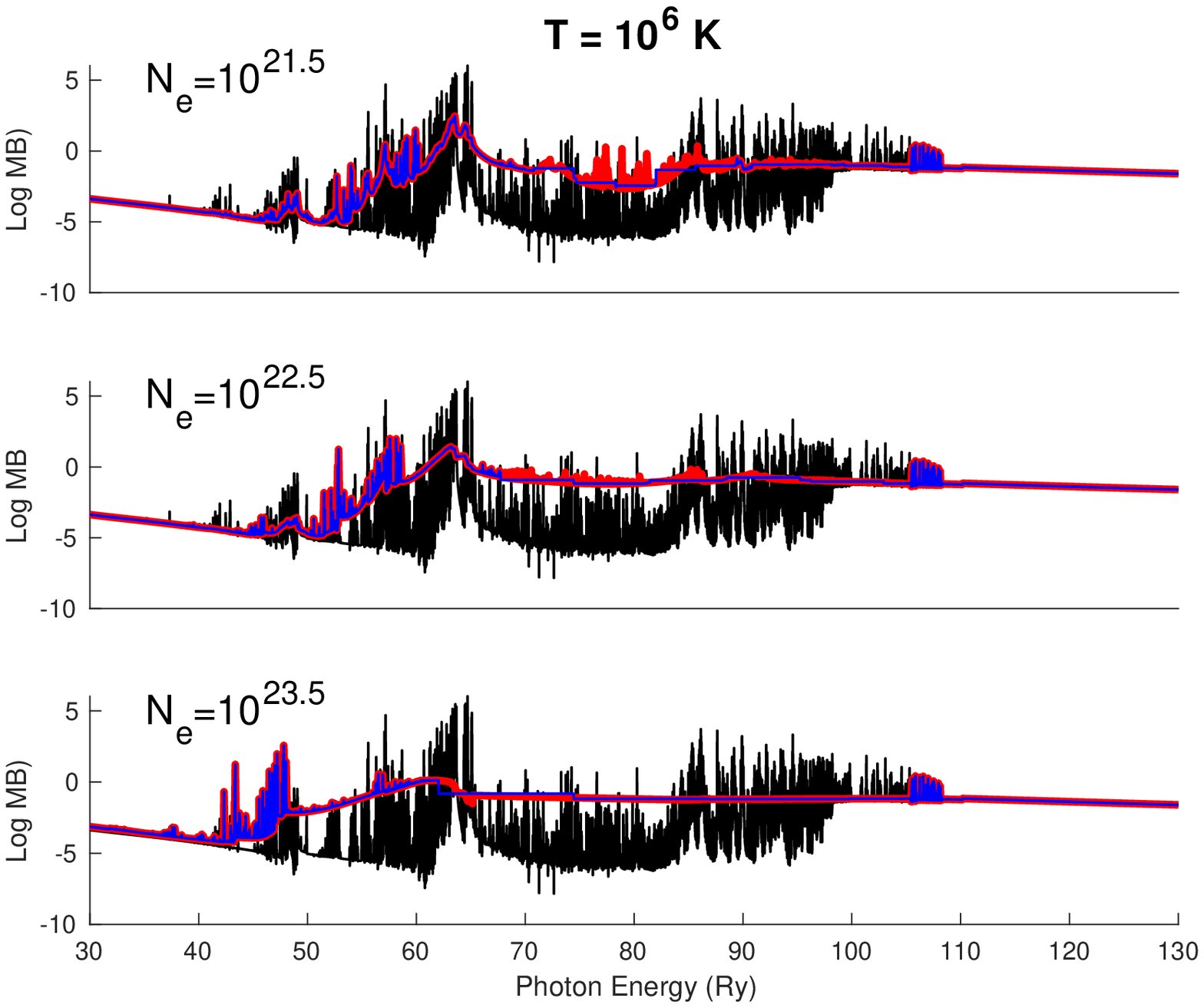}}
\end{center}
\caption{Plasma broadened photoionization cross sections
of \fexvii level $2s^22p^5[^2P^o_{3/2}]4d(^1F^o_3)$ (ionization energy
17.626 Ry), along isotherm
$T= 10^6$K and electron densities $N_e=10^{21.5,22.5,23.5}$cc, as in
Fig.~2. \label{fig:t6}}
\end{figure}

Fig.~2 displays detailed results for plasma broadened and unbroadened
photoionization cross section
of one particular excited level $2s^22p^5[^2P^o_{3/2}]4d(^1F^o_3)$
(ionization energy =
17.626 Ry) of \fexvii at three
representative densities (note the $\sim$10 orders of magnitude
variation in resonance heights along the Y-axis). The main feature
evident in the figure are as follows.
(i) AI resonances begin to show significant broadening and smearing of
a multitude of overlapping Rydberg series at
$N_e = 10^{21}$cc. The narrower high-\en \el resonances dissolve into
the
continua but stronger low-\en \el resonance retain their asymmetric
shapes with attenuated heights and widths. (ii) As the density
increases by one to two order of magnitude, to $N_e=10^{22-23}$cc,
resonance
structures not only broaden but their strengths shift and redistributed
over a wide range determined by total width
$\Gamma(\omega,\nu_i,T,N_e)$ at each energy $\hbar \omega$ (Eq. 4).
(iii) Stark ionization cut-off (Table 1) results in step-wise structures
that represent the average due to complete dissolution into continua.
(iv) The total AI resonance strengths are conserved, and
integrated values generally do not deviate by more than 1-2\%.
For example, the three cases in Fig.~2:
unbroadened structure (black), and broadened without (red) and with
Stark
cut-off (blue), the integrated numerical values are 59.11, 59.96, 59.94
respectively. This is also an important accuracy
check on numerical integration and the computational algorithm,
as well as the choice of the parameter $\delta$ that determines the
energy range of the Lorentizan profile at each T and $N_e$; in the
present
calculations it varies from $\delta$ = 0.01-0.05 for
\dne=$10^{21-24}$cc.

Fig.~3 shows
similar results ito Fig.~2 for another excited \fexvii level
$2s^22p^5[^2P^o_{3/2}]4d(^1F^o_3)$ (ionization energy
17.626 Ry), along a lower
temperature $10^6$K isotherm at different intermediate densities. Both
Figs.~2 and 3
show a redward shift of low-\en resonances and dissolution of high-\en
resonances. In addition, the background continuum is raised owing to
redistribution of resonance strengths, which merge into one across high
ying and overlapping thresholds. Free-free transitions are not
considered in the results in Figs.~2 and 3 but included in the
resutls discussed in Table 1, although it is found to be practically
negligible.

 Table 1 gives plasma parameters corresponding to Figs.~2 and 3. Their
physical significance is demonstrated by a representative sample
tabulated temperature T(K) and \dne. The maximum
width $\Gamma_{10}$ corresponding to $\nu_i=10$ in Eqs. (4-7) is set
by the CC-BPRM calculations which delineate unbroadened AI resonance
profiles up to $\nu \leq 10$, and employ an averaging procedure up to
each
threshold $ 10 < \nu < \infty$ using quantum defect (QD)
theory (viz. \cite{nb3,op95,aas} and references therein).
$\Gamma_c(10)$ and $\Gamma_s(10)$ are the maximum collisional and Stark
width components. The Doppler width $\Gamma_d$ is much smaller,
$1.18\times 10^{-3}$ and $1.67\times 10^{-3}$ Ry at $10^6$K and $2\times
10^6$K respectively, validating its inclusion in Eq.~(4) in HED plasma
sources but possibly not when $\Gamma_d$ is
comparable to $\Gamma_c$ or $\Gamma_s$. The $\nu^*_s$ and
$\nu_D$ are effective quantum numbers corresponding to Stark
ionization cut-off and the Debye radius respectively. We obtain
$\nu_D = \left[ \frac{2}{5}\pi z^2 \lambda_D^2 \right ]^{1/4}$,
where the Debye length $\lambda_D = (kT/8\pi N_e)^{1/2}$.
It is seen in Table 1 that
$\nu_D > \nu^*_s$ at the T, \dne considered, justifying neglect of
plasma screening effects herein,
but which may need to be accounted for at even higher
densities.

 The aggregate effect of AI broadening for large-scale applications
is demonstrated in Table 1 by the
ratio R of the Rosseland Mean
Opacity using broadened/unbroadened cross sections for 454 \fexvii
levels with AI resonances (other higher bound levels have negligible
resonances) \cite{nb4,np16}. For any atom or ion R
is highly dependent on T and \dne; for \fexvii R yields up to 58\%
enhancement due to plasma broadening with increasing
\dne along the $2\times 10^6$K isotherm,
but decreasing to 6\% along the $10^6$K isotherm. Approximately 70,000
free-free transitions among +ve energy levels are included in the
calculation of
R, but their contribution has no significant broadening effect since
they entail very high-lying levels with negligible level populations.
However, different plasma environments with intense radiation fields, or
a different equation-of-state than \cite{mhd} employed here, may lead to
more discernible effect due to free-free transitions. AI broadening
in a plasma environment
affects each level cross section differently, and hence
its contribution to opacities or rate equations for atomic processes
in general. A critical (T,\dne) range can therefore be numerically
ascertained where redistribution and shifts of atomic resonance
strengths would be significant and cross sections should be modified.

\begin{table}
\caption{Plasma parameters along isotherms in Fig.~2 and
3; $\nu_D$ corresponds to Debye radius; R is the ratio of \fexvii
Rosseland Mean Opacity with and without broadening
\cite{nb4}; $\Gamma_{10}$ is the maximum AI resonance width at
$\nu=10$.}
\begin{center}
\begin{tabular} {c|c|c|c|c|c|c|c}
T(K) & $ N_e (cc)$ & $\Gamma_{10}(Ry)$ & $\Gamma_c(10)$ & $\Gamma_s(10)$
&
$\nu_s^*$ & $\nu_D$ & R \\
& & $\nu=10$ & & & & &\\
hline
 $2 \times 10^6$ & $10^{21}$ & 3.42(-1)& 8.55(-2) & 2.57(-1) & 10.4 &
28.1& 1.35 \\
 $2 \times 10^6$ & $10^{22}$ & 2.05(0) & 8.55(-1) & 1.19(0) & 7.7 &
15.8 & 1.43 \\
 $2 \times 10^6$ & $10^{23}$ & 1.41(1) & 8.55(0) & 5.53(0) & 5.6 & 8.9 &
1.55 \\
 $2 \times 10^6$ & $10^{24}$ & 1.11(2) & 8.55(1) & 2.57(1) & 4.1 & 5.0 &
1.58 \\
 $10^6$ & $3.1\times10^{21.5}$ & 8.17(-1) & 2.71(-1)& 5.46(-1)& 9.0 &
17.8 & 1.47 \\
 $10^6$ & $3.1\times10^{22.5}$ & 5.25(0) & 2.71(0) & 2.53(0)& 6.6 & 10.0
& 1.13 \\
 $10^6$ & $3.1\times10^{23.5}$ & 3.89(0) & 2.71(1)& 1.18(0) & 4.8 & 5.6
& 1.06 \\
\hline
\end{tabular}
\end{center}
\end{table}

\section{Conclusion}

 The main conclusions are: (I) The method described herein is generally
applicable to AI resonances in atomic processes in HED plasmas. (II)
The cross sections become energy-temperature-density dependent in a
The cross sections become energy-temperature-density dependent in a
critical range leading to broadening, shifting, and dissolving into
continua. (III) Among the approximations
necessary to generalize the formalism is the assumption that thermal
Doppler widths are small compared to collisional and Stark widths as
herein, but given the intrinsic asymmetries of AI resonances it may not
lead to significant inaccuracies (although that needs to be verified in
future works). (IV) The treatment of Stark broadening and ionization
cut-off is {\it ad hoc}, albeit based on the equation-of-state
formulation \cite{mhd} and consistent with previous works \cite{op95}.
(V) Since it is negligiblhy small,
the free-free contribution is included post-facto
in the computation of the ratio R in Table 1 and not in the
cross sections and results shown in Figs.~2 and 3, but may be
important in special HED environments with intense radiation and should
then be incorporated in the main calculations of total AI width (Eq.~4).
(VI) The predicted redward shift of AI resonances as the plasma
density increases should be experimentally verifiable.
(VII) Redistribution of AI resonance strengths should particularly
manifest itself in rate coeffcients for
\eion excitiaton and recombination in plasma models and simulations, and
for
photoabsorption in opacity calculations, using temperature-dependent
Maxwellian, Planck, or other particle distribution functions.
(VIII) The treatment of
individual contributions to AI broadening may be improved,
and the theoretical formulation outlined
here is predicated on the assumption that external plasma effects are
perturbations subsumed by and overlying the intrinsic autoionization
effect.
(IX) The computational formalism is designed to be amenable for
practical
applications and
the computational algorithm and a general-purpose program AUTOBRO
are optimized for large-scale computations
of AI broadened atomic cross sections in HED plasma models. The CPU time
required depends mainly on the density which
determines the total width $\Gamma$; for example, in the reported
calculations
for \fexvii at T=$2\times10^6$K it is few minutes for \dne = $10^{21}$
cc and
$\sim$3 hours for \dne=$10^{24}$cc.

\vskip 0.25in
{\bf Acknowledgements}
\vskip 0.25in

 I would like to thank Sultana Nahar for \fexvii atomic data and
discussions.


\begin{thebibliography}{10}
\bibitem{nb1} Several formulations were employed to implement a Voigt
profile but were numerically unstable owing to $\Gamma_c >> \Gamma_d$.
\bibitem{nb3} The QD regions $10 < \nu_i < \infty$ are
generally small and overlapping for closely spaced excited core
levels, as herein.
\end{thebibliography}
\end{document}